\documentstyle[prb,aps,psfig,twocolumn]{revtex}
\begin{document}
\draft

\onecolumn

\title{A density functional study of lithium bulk and surfaces}
\author{K. Doll, N. M. Harrison, and V. R. Saunders}
\address{CLRC, Daresbury Laboratory, Daresbury, Warrington, WA4 4AD, UK}

\maketitle

\begin{abstract}

We report the bulk and surface properties of lithium computed
within a full potential LCGTO formalism using both
density functional theory and the Hartree-Fock approximation.
We examine the convergence of computed properties with respect
to numerical approximations and also explore the use of finite 
temperature density functional theory.
We demonstrate that fully converged calculations reproduce
cohesive properties, elastic constants, band structure, and 
surface energies in full agreement with experimental data 
and, where available, previous calculations.

\end{abstract}

\pacs{ }

\narrowtext
\section{Introduction}

Lithium has been the subject of considerable interest over many 
years. Although its electronic structure is relatively
simple, its structural properties 
still pose a significant challenge to both experiment
\cite{Barrett,Smith,Schwarz}
 and simulation 
\cite{Perdewetal,Staikov,Liuetal,Sliwko,DacorognaCohen,BoettgerTrickey1985,Nobel,Sigalas,BoettgerAlbers,Skriver,YoungRoss,Bross}.
Lithium is very soft; the determination of its elastic 
constants and surface energies requires experiments of high
accuracy and simulations of high numerical stability.
The calculation of surface formation energies is particularly
delicate as has been discussed recently
\cite{Kokkoetal,BoettgerTrickey1992}.

The aim of this article is to present a comprehensive and systematic study of
the band structure, cohesive energy, elastic constants,
phase stability, and surface energies of lithium.
Very few systematic 
studies of the dependence of results on  the computational parameters
are available. This however is especially important when energy differences
are required as for example in calculation of  surface energies or phase
stabilities.
We present the results of fully converged,
full potential, all-electron calculations based on both 
density functional theory and the Hartree-Fock approximation. 
We examine
the use of finite temperature density functional theory as
a technique for accelerating convergence with respect to
reciprocal space sampling. We expand the crystalline orbitals as 
a linear combination of Gaussian type orbitals (LCGTO). This 
approach is very well established for insulators
\cite{CRYSTALbuch,PisaniBuch}. We find, in accord with
a recent study of magnesium \cite{Baraille}, that this
approach is also well suited to the simulation of
a free-electron metal.

The paper is organized as follows: in section \ref{section2}, we discuss
the computational parameters. In sections \ref{section3} and \ref{section4},
we discuss results for lithium bulk and surfaces, respectively, and
summarize our conclusions in section \ref{section5}.

\section{Basis set and Method}
\label{section2}
All the calculations were performed with the program package
{\sc crystal} \cite{Manual}.
The main numerical approximation in our approach
is the choice of the Gaussian basis set. The difficulties
of selecting basis sets for metallic systems have been
explored in previous studies \cite{Dovesi1982,Dovesi1983}.
In principle the quality of a calculation can be systematically
improved by adding additional functions to the basis set
and optimizing their exponents in a suitable reference 
state - usually the bulk crystal. In practice one must balance
the overcompleteness of the basis set, which leads to 
linear dependence, with the need for additional variation
freedom. For molecular systems and insulating solids
these problems have largely been overcome and Gaussian 
basis sets are very widely used. For metallic systems
and their surfaces in particular there have been 
very few systematic studies.

In a solid the tails of the atom centered functions overlap
strongly and so diffuse basis functions optimized for the
description of atomic or molecular systems are not useful
and indeed may give rise to linear dependence. We are
thus unable to simply use basis sets from the many libraries
developed for the description of molecular systems.
We have therefore developed
an hierarchy of basis sets of increasing quality in order
to examine the convergence of computed properties.

The smallest basis set used has 3 $s$-symmetry functions and 
2 $p$-symmetry functions and is denoted as $[3s2p]$ . The $[1s]$ 
radial function was taken from Ref. \onlinecite{Prencipe}. 
The exponents of the two additional $sp$-shells
were optimized in local density approximation (LDA) 
calculations 
(with Dirac-Slater exchange \cite{DiracSlater}
 and Perdew-Zunger's correlation functional
\cite{PerdewZunger})
 for the solid at the experimental lattice constant. 
The lowest energy was obtained with exponents of 0.50 and 0.08. Similar
results were obtained when using the 
Perdew Wang gradient corrected approximation 
(PWGGA)\cite{Perdewetal}. However,
as an exponent of 0.08 gives rise to a very diffuse basis function
close to numerical instability,
instead exponents of 0.50 and 0.10 were chosen. This $[3s2p]$
basis set is very robust and computationally efficient --- it does
not give rise to linear dependence even
when the bulk is strongly distorted (for example, 
to determine the elastic constants).

A $[4s3p]$ basis set was obtained by using three exponents
(0.50, 0.20, and 0.08) - which were chosen to be "even tempered", 
i.e. the ratio between the exponents is kept fixed (2.5 in this case). 
This ratio is close to the lowest which can be tolerated before
on-site (atomic) linear dependence is seen. It is however also
known to converge the atomic energy to within less than $10^{-4} E_h$ 
 ($E_h$ = 27.2114 eV)
of the
exact Hartree-Fock ground state energy (see the analysis in Ref. 
\onlinecite{SchmidtRuedenberg}). Finally, an additional polarization
function of $d$-symmetry was added and the exponent optimized
within a PWGGA calculation to be  0.15. 
However, the $d$-function leads only to a minor change in
the total energy. The energy varies only by $5 \times 10^{-5} E_h$ 
when, e.g., changing
the exponent to 0.5. 
The basis sets developed in this manner are displayed in table
\ref{CRYSTALbasis}.

Both at the Hartree-Fock (HF) and B3LYP  \cite{B3LYP}
(invoving a hybrid of Fock exchange and a modification of
the Becke gradient corrected exchange functional\cite{B3,B3PW}, and
the Vosko-Wilk-Nusair local correlation functional V 
\cite{VWN}  
and the gradient corrected correlation potential by Lee, Yang and Parr
\cite{LYP})
levels, an optimization of basis set exponents was not possible. Instead,
the outermost exponent became more and more diffuse until finally
the solution became unstable. This is a well known pathology of
the use of Fock exchange in metallic systems 
(see also the discussion in Ref. \onlinecite{Dovesi1983}). 
When features of the HF solution are discussed in
this article,
they were obtained with the $[3s2p]$ basis set 
(outermost exponents 0.50 and 0.10).

In order to compute binding energies the free atom is calculated 
within a spin-polarized formalism with the same $[1s]$ function
but with additional $s$-exponents 0.60, 0.24, 0.096, 0.04, and 0.016
to describe the long range behaviour of the atomic wavefunction.

For the LDA and PWGGA calculations we also expand the exchange and 
correlation potentials in an auxiliary Gaussian basis set which consists
of 13 even tempered $s$-functions with 
exponents from 0.1 to 2000, 3 even tempered $p$-functions with
exponents from 0.1 to 0.8, and 2 $d$-functions with exponents of 0.12 and 0.3.
This is sufficient to integrate the charge density to an accuracy 
of 10$^{-7}$ $ |e|$. For the free atom, we use an auxiliary basis set
with 18 even tempered $s$-functions with exponents from 0.0037 to 4565.

Reciprocal space sampling is a delicate problem especially in metals.
The sampling is performed on a Pack-Monkhorst net \cite{PackMonkhorst}
where the density of points is determined by a shrinking factor.
The Fermi energy and shape of the Fermi surface are determined by
interpolation onto a "Gilat" net. This net is simply related to
the Pack-Monkhorst net by an additional subdivision factor.
To further improve convergence, the finite temperature generalization of
density functional theory\cite{Mermin} can be used to apply Fermi surface
smearing \cite{FuHo}.
In table \ref{kpointconvergencetable}, the dependence of the total
energy on the density of points in the Pack-Monkhorst net
 is displayed.
For this purpose, PWGGA calculations on a body centered cubic (bcc) lattice 
at the equilibrium lattice constant of 3.44 \AA \mbox{ } were performed.

\begin{figure}
\centerline{\psfig{figure=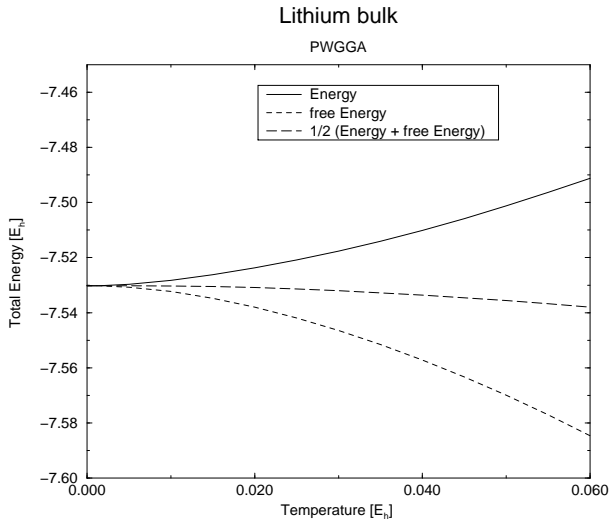,width=9cm,angle=270}}
\caption{Energy $E(T)$, free energy $F(T)$ 
and $\frac{1}{2}(E(T)+F(T))$ of Li bulk, bcc lattice. A shrinking factor
of 16 was used.}
\label{EFE0bulkfigure} 
\end{figure}

At zero temperature, even with the largest net used,
the energy is  still slightly decreasing as more points are used.
For a smearing of 0.001 $E_h$ 
(which corresponds to a temperature of $T=
\frac{0.001 E_h}{k_B}= 316 K$ with Boltzmann's constant $k_B=3.1667 \times 
10^{-6} \frac{E_h}{K}$), the energy is stable up to  a few
$\mu E_h$. A higher number of sampling points in the Gilat net leads to a
systematic improvement at zero temperature. At finite temperature,
the number of sampling points in the Gilat net does not influence the
results when a sufficiently high number in the Pack-Monkhorst net
is chosen. As shown in table \ref{kpointconvergencetable},
the difference in energy for the different number of sampling points
in the Gilat net is of the order of only a few $\mu E_h$ 
for a fixed shrinking factor of 24 in the Pack-Monkhorst net.
Note that for the most dense net at $k_BT=0.001 E_h$, the difference
in energy between smeared and unsmeared results is less that $10^{-4} E_h$.
At an even higher temperature of 0.02 $E_h$, the energy converges to
a value which is 6 $mE_h$ higher than at 0.001 $E_h$. An estimate of the
zero temperature energy is possible by using the approximation
$E(0 K)=\frac{1}{2}(E(T)+F(T))$ (with 
$F=E-TS$ being the free energy and exploiting the fact that the energy 
increases quadratically with temperature for low temperature)
as suggested in Ref. \onlinecite{Gillan}. The electronic 
entropy $S$ is defined as\\
$S=k_{B} \sum_{i}^{N_{states}}  f_i \ln f_i  + (1-f_i)\ln(1-f_i)$ \\
with $f_i$
being the Fermi function.
This leads to a value of $E(0)$ extrapolated from $k_B T=0.02 E_h$
 which deviates by less than $10^{-3} E_h$ from
the value obtained at a temperature of 0.001 $E_h$.
The functions $E(T)$, $F(T)$ and $\frac{1}{2}(E(T)+F(T))$ 
for a fixed value of 145 sampling points (corresponding to
a shrinking factor of 16 in 
the Pack-Monkhorst net) are also displayed
in Figure \ref{EFE0bulkfigure}. Indeed, even for relatively high temperature,
$E(0)$ is well approximated by  $\frac{1}{2}(E(T)+F(T))$.

In conclusion, as we need a high accuracy for the energy difference
between the different phases of Li,
we chose for the calculations on lithium bulk a shrinking factor of 16
for the Pack-Monkhorst net which gives 145  sampling points in the
irreducible Brillouin zone of the non-distorted bulk
and a temperature
of 0.001 $E_h$. 
This ensures that convergence of the energy to at least $10^{-4} E_h$
with respect to  reciprocal space sampling is achieved.

\section{Results for bulk Lithium}
\label{section3}
\subsection{Band structure}

\begin{figure}
\centerline{\psfig{figure=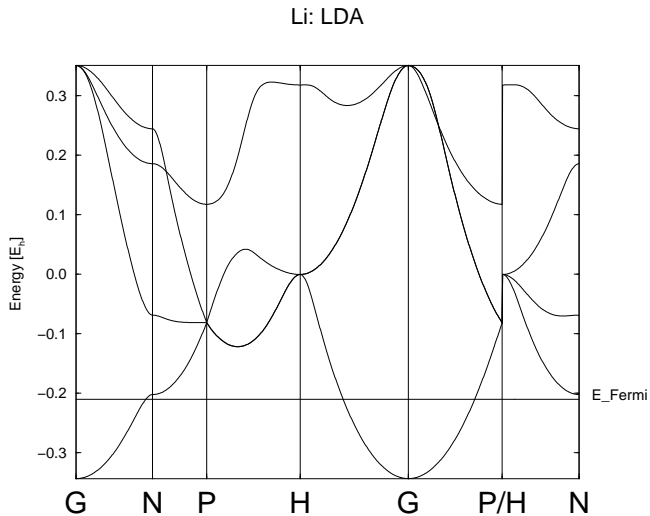,width=9cm,angle=270}}
\caption{LDA band structure at the equilibrium lattice constant, bcc lattice}
\label{LDAband} 
\end{figure}
In figure \ref{LDAband} 
the LDA band structure
for the bcc structure
is displayed.
The occupied bands and the lower unoccupied bands are 
in excellent agreement with results earlier obtained with Gaussian basis sets
and $X_{\alpha}$ exchange \cite{ChingCallaway}, by the 
Kohn-Korringa-Rostoker (KKR)
method \cite{Lawrence} and Slater exchange, 
augmented plane wave (APW) 
\cite{Papaconstantopoulos} and modified APW (MAPW)
\cite{Bross} calculations both using LDA,
or linear muffin tin orbital (LMTO) calculations \cite{KotaniAkai} with
a combination of exchange using the Langreth and Mehl functional 
\cite{LangrethMehl} and LDA correlation.
When using PWGGA instead of LDA, the band structure does 
not exhibit major differences. The experimentally known
conduction band width ($\sim$ 4 eV) \cite{CrispWilliams} 
is slightly lower than the result
calculated here (4.6 eV). The slow decay of the density matrix in metals
leads to difficulties when Fock exchange is involved: the
summation of the exchange series in direct space is very long 
ranged and is truncated at a large but finite distance.
This cutoff for large distances in direct space
results in numerical instabilities for small $\vec k$. Thus,
both in the Hartree-Fock and B3LYP band structures,
artificial oscillations can be found around the Gamma point.
As usual for Hartree-Fock calculations, we find that the bandwidth
is roughly twice as large as the experimental bandwidth.

\subsection{Cohesive properties}

Table \ref{groundstatetable} 
gives results for ground state properties of bcc Li
(lattice constant, bulk modulus, cohesive energy and elastic constants
C$_{11}$, $C_{11}-C_{12}$, and C$_{44}$). The elastic constants were
obtained by applying a rhombohedral distortion for $C_{44}$,
a tetragonal distortion for $C_{11}$, and
an orthorhombic distortion for $C_{11}-C_{12}$ to  
the solid at the equilibrium lattice constant.
Cohesive energy and lattice constant agree well with experiment and depend
only weakly on the basis set.  The bulk modulus is
obtained from the energy as a function of volume and agrees within the
accuracy of the fit with that obtained using the 
relation $B=\frac{1}{3}(C_{11}+2C_{12})$.
Elastic constants have a strong dependence on the
basis set and the deviation from experiment improves especially when
going from $[3s2p]$ to $[4s3p]$; the $d$-function
has only minor impact on the results.
We find that
PWGGA is closer to experiment with results similar to Ref. 
\onlinecite{Perdewetal}.

\subsection{Relative stabilities}

\begin{figure}
\centerline
{\psfig{figure=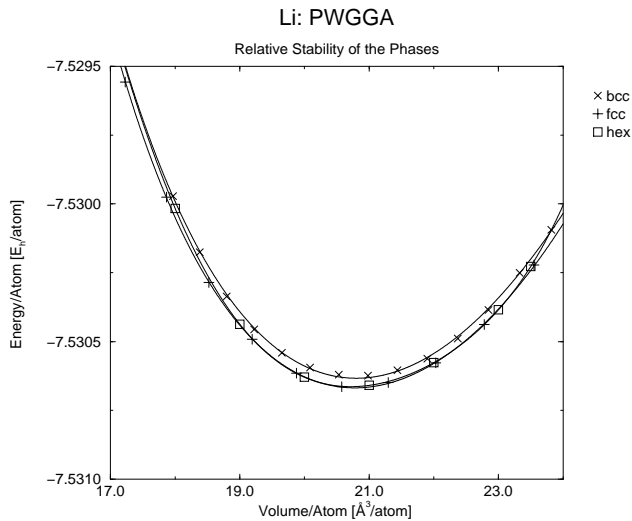,width=9cm,angle=270}}
\caption{Relative stabilities of the different phases as a function of 
volume, $[4s3p1d]$ basis.}
\label{relstabbild}
\end{figure}

The experimental 
crystal structure of lithium at zero temperature is still unclear
(at room temperature a bcc structure is favoured). Both 
face-centered cubic (fcc) and hexagonal (hex) structures have been suggested
as well as more sophisticated structures such as 9R (a nine layer
sequence of close-packed planes ABCBCACAB) \cite{Overhauser} 
or mixtures of these phases. Therefore, we also
investigated the relative stability of the different phases.
In Figure \ref{relstabbild}, 
total energies of bcc, fcc and hex phase, obtained with the
PWGGA functional and the best basis set ($[4s3p1d]$), are displayed.

The c/a ratio
of the hexagonal phase remains close to the ideal close packed
value of 1.633 (it varies between 1.631 and 1.635 which is within
the accuracy of the fit).
We find that the closed-packed structures are slightly lower in energy
than the bcc structure. Although it is reasonable to conclude that
the close packed structures are lower in energy than bcc it is not
possible to resolve the difference in energy between the fcc and hex
phases. The variation of the energy with basis set is given in
table \ref{relstabtable}. As the basis set is systematically improved
the energy difference between bcc and fcc increases from 
1 $\times 10^{-5} E_h$  to 4 $\times 10^{-5} E_h$,  and between bcc and hex
decreases from 7 $\times 10^{-5} E_h$  to 
4 $\times 10^{-5} E_h$. We note that even the $d$-function influences 
this energy splitting.
These results, which do not include the zero point energy,
indicate a preference for closed-packed
structures which is in agreement with most of the previous calculations
\cite{Perdewetal,Staikov,Liuetal,Sliwko,DacorognaCohen,BoettgerTrickey1985,Nobel,Sigalas,BoettgerAlbers,Skriver,YoungRoss} except for one \cite{Bross} 
(see table 
\ref{relstabtable}). 

HF calculations were only possible with the smallest $[3s2p]$ basis set
where the order of phases is different with hcp being the lowest in energy,
followed by bcc and fcc being highest. The same is found in LDA with
the smallest $[3s2p]$ basis set, but changes when the basis set
is increased to $[4s3p]$ where both fcc and hcp are  2 
$\times 10^{-4} E_h$
lower in energy than bcc (again, calculations with the best $[4s3p1d]$
basis set were not possible because of linear dependence) ---
table \ref{relstabtable}.

As the energy splitting between the close packed and bcc phases
is rather small the zero-point energy difference cannot be neglected. 
The first published calculation of this, based on the harmonic approximation,
\cite{Staikov} gave an additional stabilization of
$9 \times 10^{-5} E_h$ of the fcc phase compared to the bcc phase.
In these calculations however the hcp phase was found to be 
higher in energy that the bcc phase (table \ref{relstabtable}).
Very recently, in a calculation also including anharmonic effects,
the stabilization was calculated to
be about $1.5  \times 10^{-5} E_h$ both for fcc and 9R phases relative to
the bcc phase \cite{Liuetal}. In addition, the authors computed the
variation of the vibrational free energy as a function of 
temperature and found a phase
transition from a closed packed to a bcc phase at a temperature
of $T \sim 200 K$.

\section{Results for surfaces}
\label{section4}

A further test of this approach is the calculation of surface energies.
We model lithium surfaces by using a slab and varying the
numbers of layers (a bcc structure is assumed). 
Surface energies can be calculated in two ways, either by 
deriving a bulk energy by subtracting energies of two slabs with 
$n$ and $m$ layers:

\begin{equation}
\label{nn-1gleichung}
E_{surface}=
\frac{1}{2}(E_{slab}(n)-(E_{slab}(n)-E_{slab}(n-m))\frac{n}{m})
\end{equation}

\noindent
which has the advantage of a systematic error cancellation (in particular
the reciprocal space sampling is consistent between the bulk and slab
energies) or by
using an independent bulk energy

\begin{equation}
\label{eebulkgleichung}
E_{surface}=\frac{1}{2}(E_{slab}(n)-E_{bulk} \times n)
\end{equation}

\noindent
All the quantities $E_{surface}$, $E_{slab}(n)$, and $E_{bulk}$ 
are energies per atom. 

\begin{figure}
\centerline{\psfig{figure=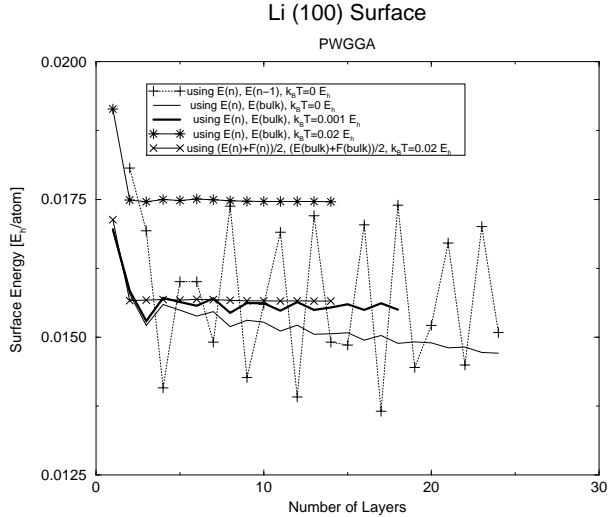,width=9cm,angle=270}}
\caption{(100) Lithium surface energy 
with a shrinking factor of 24 in the Pack-Monkhorst net,
$[3s2p]$ basis.}
\label{100k24-48alles}
\end{figure}

In Figure \ref{100k24-48alles}, 
results for surface energies using both equation
\ref{nn-1gleichung} (with $m$=1) and \ref{eebulkgleichung} are displayed.
Equation \ref{nn-1gleichung} leads to relatively strong oscillations
(dotted line with pluses)
 and gets
more stable the larger $m$.
Numerical noise in the expression
 $(E_{slab}(n)-E_{slab}(n-m))\frac{n}{m}$ is reduced for larger values of $m$.
Equation \ref{eebulkgleichung}, however, leads at zero temperature
to a slight linear decreasing  of the
surface energy as a function of the number of layers (thin line without
additional symbol).
The reason for the nonvanishing
slope is that
the energy difference $E(n)-E(n-1)$ is not identical to the energy
of the bulk due to the systematic errors in the convergence of the
total energy with respect to reciprocal lattice sampling
(this was also emphasized in Refs. 
\onlinecite{Kokkoetal,BoettgerTrickey1992}).
As shown in table \ref{kpointconvergencetable}, at zero temperature
the bulk energy is still changing on the order of $10^{-4} E_h$  with
increasing sampling point number. Similarly the bulk energy varies
when extracted from the slab. 
This slight discrepancy gives rise to a variation of the surface energy
with the number of layers with a slope of 
$10^{-4} E_h/atom/layer$.

The origin of the poor convergence with respect to reciprocal lattice
sampling is due to the sharp cutoff imposed by the Fermi energy.
One possibility to obtain the surface energy would be to use the intercept
from a linear fit, but
a better and simpler way to alleviate the difficulties associated with
reciprocal lattice sampling is smearing the Fermi surface with a finite
temperature. Already at a temperature
of 0.001 $E_h$, the slope is virtually zero (thick line without additional
symbol).
This is consistent with table  \ref{kpointconvergencetable} as the
bulk energy converges much faster at finite temperature.
At a  higher temperature of $k_BT= 0.02 E_h$, 
$E(T)$ clearly deviates from $E(0)$ (thin line
with stars) and
the approximation $E(0)=\frac{1}{2}(E(T)+F(T))$ should be applied.
This works very well when comparing the corresponding results
(thin line with crosses) with
results calculated at $k_BT=0.001$ $E_h$ (thick line without additional
symbol). The surface energy obtained this way is only slightly higher
than that from a calculation at $k_BT=0.001$ $E_h$  which is consistent
with figure \ref{EFE0bulkfigure}.

\begin{figure}
\centerline{\psfig{figure=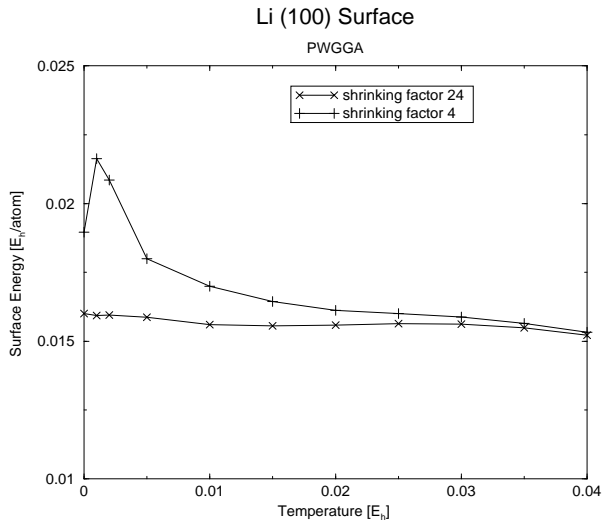,width=9cm,angle=270}}
\caption{(100) Lithium surface energy with two different reciprocal lattice
samplings extracted from two slabs with 5 and
6 layers using $\frac{1}{2}(E(T)+F(T))$ and Equation \ref{nn-1gleichung}}
\label{100surfaceK24-48K4-4tempfigure}
\end{figure}

A higher smearing temperature also reduces the oscillations 
in the curves both when using equation \ref{nn-1gleichung} 
or \ref{eebulkgleichung}.
It even allows to substantially reduce the
number of sampling points as shown in Figure 
\ref{100surfaceK24-48K4-4tempfigure}. At low temperature, a high number
of sampling points is necessary to obtain the correct result whereas
at high temperature already a shrinking factor of 4 (resulting in 
6 sampling points in the irreducible zone) is sufficient. It should
be noted that in this case we used Equation \ref{nn-1gleichung} so that all 
the data is consistently extracted from calculations on slabs.

In conclusion, calculations at zero temperature are very cumbersome
when results from calculations on slabs and on the bulk have to be combined. 
The error cancellation can be maximized by extracting the
surface energy from calculations on slabs only. The 
problem of error cancellation  between bulk and slab
is already improved at  very low temperature when it is possible
to fully converge the bulk energy with respect to reciprocal lattice
sampling.
Higher temperature also leads to a smoother behaviour of the
surface energy as a function of number of layers. Finally,
calculations at high temperature can be performed with a strong 
reduction of the number of sampling points and
using $E(0)=\frac{1}{2}(E(T)+F(T))$
as an approximation for zero temperature results as suggested in Ref. 
\onlinecite{Gillan}.

In table \ref{surfsummtable}, LDA and PWGGA results for the unrelaxed
(100), (110) and (111)
surface are summarized. The lattice constant was chosen as the
 bulk equilibrium lattice constant, 
a temperature of 0.001 $E_h$ and
a higher shrinking factor of 24 
resulting in 91 sampling points for the slabs and 413 sampling points
for the bulk was used. 

 As found in Ref. 
\onlinecite{Perdewetal} for the
jellium model without additional long-range corrections,
surface energies in PWGGA are lower than
in LDA 
(the different lattice constants for PWGGA and LDA are not the reason,
an evaluation at the LDA equilibrium lattice constant leads to a 
change of the PWGGA surface energy which is negligible compared to the 
difference between LDA and PWGGA surface energies). The energies
are reduced when going from the smaller $[3s2p]$ to the $[4s3p]$ basis
set, introducing a $d$ function leads only to minor changes.
Our results  agree well with the literature 
\cite{Kokkoetal,SkriverRosengaard}.

\section{Conclusion}
\label{section5}

In this study of lithium metal, we presented accurate 
results using a full potential, all electron
density functional scheme.  Results for cohesive
properties, elastic constants, band structure, 
and surface energies are in full agreement with experiment and 
calculated values from literature. The results 
in best agreement with experiment
were obtained with the
 gradient corrected functional of Perdew and Wang. Hartree-Fock
calculations for lithium are very difficult as it is impossible to optimize
the exponents because of the very long range
of the exchange interaction; the same problems appear in functionals
involving an admixture of Fock exchange.
We demonstrated the convergence of the different properties with
respect to the computational parameters by using an hierarchy of basis sets,
different reciprocal lattice samplings and different smearing temperatures.
We showed that finite temperature calculations can be used to improve
convergence and still an extrapolation to zero temperature is possible
and accurate.
Quantities like cohesive energy and lattice constant are already 
stable with the smallest basis set; elastic constants 
and surface energies are more sensitive. The most difficult quantity
to calculate is the energy
splitting between the different phases where we have reached the
limit of numerical accuracy. We
can not make a prediction about the preferred crystal structure; 
the energy difference is so small that, from the computational point
of view, even subtle changes such as introducing a $d$-function are important
and
zero point energies must be included \cite{Staikov,Liuetal}.
We confirm the finding of Ref. \onlinecite{Baraille}
that an approach based on Gaussian type functions provides a reliable
and very efficient description of metallic systems.

\onecolumn

\begin{table}
\begin{center}
\caption{\label{CRYSTALbasis}Basis sets}
\vspace{5mm}
\begin{tabular}{ccccc} 
& exponent & $s$ contraction & $p$ contraction & $d$ contraction\\
$[1s]$    &  840.0  &  0.00264 \\
          & 217.5   & 0.00850\\
          &  72.3   & 0.0335\\
          &  19.66  & 0.1824\\
          &   5.044 & 0.6379\\
          &   1.5   & 1.0\\
\\
$[3s2p]$\\
$[2sp]$ & 0.50 & 1.0 & 1.0\\
$[3sp]$ & 0.10 & 1.0 & 1.0\\
\\
$[4s3p]$ and $[4s3p1d]$\\
$[2sp]$ & 0.50 & 1.0 & 1.0\\
$[3sp]$ & 0.20 & 1.0 & 1.0\\
$[4sp]$ & 0.08 & 1.0 & 1.0\\
$[d]$ & 0.15 &  & & 1.0\\
\end{tabular}
\end{center}
\end{table}

\begin{table}
\begin{center}
\caption{Convergence of the total energy with respect to the number of 
sampling points. The results are obtained from PWGGA calculations
with a $[3s2p]$ basis set, for the  
bcc lattice at a lattice constant of 3.44 \AA and for 
three temperatures 0 $E_h$, 0.001 $E_h$, 
and 0.02 $E_h$.}
\label{kpointconvergencetable}
\vspace{5mm}
\begin{tabular}{ccccccc}
 & &  &  & \\
shrinking & number of & \multicolumn{4}{c}{Shrinking factor of Gilat net } \\
factor of the & sampling points in 
& \multicolumn{4}{c}{in multiples of shrinking factor}\\
Pack Monkhorst & irreducible part of the & \multicolumn{4}{c}{of the
Pack-Monkhorst net}\\
 net & Pack-Monkhorst net &\\
\\
$k_{B}T=0 E_h; E(T)=E(0)$\\
  & & 1 & 2 & 3 & 4 \\
4 & 8 & -7.520822 & -7.520213 & -7.520284 & -7.520373 \\
8 & 29 & -7.527049 & -7.529923 & -7.530344 & -7.530488 \\
12 & 72 & -7.528987 & -7.529867 & -7.530023 & -7.530075  \\
16 & 145 & -7.529574 & -7.530032 & -7.530125 \\
18 & 195 &  -7.529697 &  -7.530077 \\
20 & 256 &  -7.529796 & -7.530101 \\
24 & 413 &  -7.529915 & -7.530130 \\
\\
$k_{B}T=0.001 E_h; E(T)$\\
\\
4 & 8 & -7.524523 & -7.522697 &  -7.521528 & -7.521029 \\
8 & 29 & -7.529710  & -7.530632 & -7.530634 & -7.530654 \\
12 & 72  & -7.529982 & -7.530156 & -7.530141 & -7.530128 \\
16 & 145 & -7.530175 & -7.530183 & -7.530181 \\
18 & 195 & -7.530195 & -7.530189 \\
20 & 256 &  -7.530185 & -7.530188 \\ 
24 & 413 &  -7.530200 & -7.530185 \\
\\
$k_{B}T=0.001 E_h; E(0)=\frac{1}{2}(E(T)+F(T))$\\
\\
4 & 8 &     -7.524782 & -7.522773 & -7.521551 & -7.521064 \\
8 & 29 &    -7.529759 & -7.530653 & -7.530658 & -7.530674 \\
12 & 72 &   -7.530024 & -7.530175 & -7.530161 & -7.530150 \\
16 & 145 &  -7.530203 & -7.530204 & -7.530201 \\
18 & 195 &  -7.530218 & -7.530211 \\                                   
20 & 256 &  -7.530208 & -7.530210 \\
24 & 413 &  -7.530220 & -7.530205 \\ 
\\
$k_{B}T=0.02 E_h; E(T)$\\
\\
4 & 8 &      -7.521660 & -7.516254 & -7.515090 & -7.514714 \\
8 & 29 &     -7.523669 & -7.524181 & -7.524207 & -7.524218 \\
12 & 72 &    -7.523698 & -7.523670 & -7.523664 & -7.523661 \\
16 & 145 &   -7.523697 & -7.523689 & -7.523687 \\
18 & 195 &   -7.523697 & -7.523701 \\                                     
20 & 256 &   -7.523697 & -7.523701 \\
24 & 413 &   -7.523697 & -7.523698\\
\\
$k_{B}T=0.02 E_h; E(0)=\frac{1}{2}(E(T)+F(T))$\\
\\
4 & 8 &      -7.529537 & -7.524058 & -7.522959 & -7.522615 \\
8 & 29 &     -7.530824 & -7.531214 & -7.531234 & -7.531242 \\ 
12 & 72 &    -7.530840 & -7.530824 & -7.530820 & -7.530818 \\
16 & 145 &   -7.530840 & -7.530833 & -7.530831 \\
18 & 195 &   -7.530840 & -7.530842 \\                                     
20 & 256 &   -7.530840 & -7.530843 \\
24 & 413 &   -7.530840 & -7.530840 \\ 
\end{tabular}
\end{center}
\end{table}

\begin{table}
\begin{center}
\caption{\label{groundstatetable}Ground state properties of lithium. Energies
are in $E_h$,
lattice constants in \AA, elastic constants in GPa.}
\vspace{5mm}
\begin{tabular}{ccccccc}
 & &  &  & \\
 & $a$ & $E_{coh}$  & $B$ & C$_{44}$ & C$_{11}$ 
& $C_{11}-C_{12}$\\
LDA ([3s2p]) 
 & 3.37 & 0.066  & 16 & 13 &  18 &  3.2 \\
LDA ([4s3p]) & 3.40 & 0.066 &  & &  & \\
PWGGA ([3s2p])
 & 3.44 & 0.059  & 14 & 12 &  17 &  3.4 \\
PWGGA ([4s3p]) & 3.46 & 0.059  & 12 & 
12 & 13 & 3.5 \\
PWGGA ([4s3p1d]) 
& 3.46 & 0.059   & 11 & 12 & 13 & 3.3 \\
HF & 3.65 & 0.020  & 12 & & &  \\
\\
Lit. (LDA ; PWGGA, FP-LAPW) \cite{Perdewetal} & 3.37 ;  3.44 & & 15.0 ; 13.4 \\
Lit. (LDA, FP-LAPW) \cite{Sliwko} & 3.36 &  & 15.4 & & 17 & 2.3 \\
Lit. (LDA, Gaussian basis) \cite{CallawayZouBagayoko} & 3.45 & 0.062 & 13.8 \\
Lit. (LDA, plane wave basis) \cite{Staikov} & 3.44 &  & 13.5 \\ 
Lit. (LDA, plane wave basis) \cite{DacorognaCohen} & 3.40 & & 13.0 \\
Lit. (LDA, Gaussian basis, KSG model) 
\cite{BoettgerTrickey1985} & 3.49 & 0.044 & 14.7 \\
Lit. (LDA, Gaussian basis, HL model) \cite{BoettgerTrickey1985}
& 3.36 & 0.062 & - \\
Lit. (LDA, Gaussian basis, RSK model) \cite{BoettgerTrickey1985} & 3.34 & 0.068 & 15.8 \\
Lit. (LDA, HL) \cite{JanakMoruzziWilliams} & 3.39 & 0.061 & 14.8 \\
Lit. (MAPW) \cite{Bross} & 3.33 & 0.074 & 15.6 \\
Lit. (HF) \cite{Dovesi1983} & & 0.010 (estimated HF limit: 0.019) \\
Lit. (HF) \cite{YaoXuWang} & 3.65 & 0.022 & 14   \\
Lit. (QMC) \cite{YaoXuWang} & 3.56 & 0.058 & 13   \\
Lit. (local Ansatz) \cite{Stollhoff} & 3.56 & 0.057 & 12 \\
exp. & 3.48 \cite{AndersonSwenson} 
& 0.061 \cite{Gschneider}  & 13.0 \cite{Felice} & 11.6 \cite{Felice} & 14.5 \cite{Felice} & 2.4 
\cite{Felice} \\
\end{tabular}
\end{center}
Acronyms are: \\
Full-potential linear augmented plane wave (FP-LAPW)\\
Rajaporal-Singhal-Kimball (RSK)\\
Kohn-Sham-Gaspar (KSG)\\
Hedin-Lundquist (HL)\\
Quantum Monte-Carlo (QMC)\\
\end{table}

\begin{table}
\begin{center}
\caption{\label{relstabtable}Relative stability of the different phases
of lithium. Energies
are in $mE_h$ units per atom relative to the bcc phase, 
lattice constants in \AA, bulk moduli in GPa.
In the 9R and hexagonal structures the lattice constant $a$ refers to the
nearest neighbour distance in the basal plane.}
\vspace{5mm}
\begin{tabular}{ccccccc}
 & &  &  & \\
\hline
phase & $a$ & $E_{bcc}-E$  & $B$ 
\\
fcc, HF ([3s2p]) & 4.59 & -0.08$^a$ & 12 \\
hcp, HF ([3s2p]) & 3.24 & 0.18$^a$ & 12 \\
fcc, LDA ([3s2p]) & 4.25 & -0.02$^a$ & 16 \\
fcc, LDA ([4s3p]) & 4.24 & 0.19$^a$ &  \\
hcp, LDA ([3s2p]) & 3.00 & 0.08$^a$ & 16 \\
hcp, LDA ([4s3p]) & 3.00 & 0.24$^a$ &  \\
fcc, PWGGA ([3s2p]) & 4.33 &   0.01$^a$    & 14 \\
fcc, PWGGA ([4s3p]) & 4.35 &   0.04$^a$      & 12 \\
fcc, PWGGA ([4s3p1d]) & 4.36 &  0.04$^a$   & 11 \\
hex, PWGGA ([3s2p]) & 3.06  & 0.07$^a$     & 14 \\
hex, PWGGA ([4s3p]) & 3.08  & 0.06$^a$        & 12 \\
hex, PWGGA ([4s3p1d]) & 3.08 & 0.04$^a$    & 11 \\
\\
fcc, Ref.  \onlinecite{Perdewetal}, LDA; PWGGA  & 4.23; 4.33 & 0.15; 0.14  
& 15.5;  13.3 \\
fcc, Ref. \onlinecite{Liuetal}, LDA & 4.28 ; 4.32$^b$ &  0.073 ; 0.057$^b$ & \\
hcp, Ref. \onlinecite{Liuetal}, LDA & 3.03  & 0.062 & \\
9R, Ref. \onlinecite{Liuetal}, LDA & 3.03 ; 3.06$^b$ & 0.065 ; 0.050$^b$ \\
fcc, Ref. \onlinecite{Sliwko}, FP-LAPW & 4.23   & 0.08 & 15.2  \\
fcc, Ref. \onlinecite{Staikov}, LDA & 4.34 &  0.09 ; 0.18$^b$ & 13.4\\
hcp, Ref. \onlinecite{Staikov}, LDA & 3.09  & -0.01 & 13.3\\
9R,  Ref. \onlinecite{Staikov}, LDA & 3.07 & 0.02 & 13.3 \\
fcc, Ref. \onlinecite{DacorognaCohen}, LDA & 4.28 & 0.1  & 13.8 \\
hex, Ref. \onlinecite{DacorognaCohen}, LDA & 3.02  &  0.33    & 13.7 \\
fcc, Ref. \onlinecite{BoettgerTrickey1985}, KSG model & 4.38 & 0.25 & 18.7 \\
fcc, Ref. \onlinecite{BoettgerTrickey1985}, RSK model & 4.20 & 0.45 & 16.8 \\
fcc, Ref. \onlinecite{Nobel}, FP-LAPW &  & 0.12 & \\
hcp, Ref. \onlinecite{Nobel}, FP-LAPW &  & 0.16 & \\
fcc, Ref. \onlinecite{Sigalas}, APW & 4.21 & 1.4 & \\
fcc, Ref. \onlinecite{Sigalas}, FP-LAPW & 4.24 & 0.24 & \\
fcc, Ref. \onlinecite{BoettgerAlbers}, LMTO & & 0.12 & \\
hcp, Ref. \onlinecite{BoettgerAlbers}, LMTO & & 0.15 & \\
fcc, Ref. \onlinecite{Bross}, MAPW & 4.21 & -0.13 & 12.0 \\
\end{tabular}
\end{center}
$^a$ {As explained in the text, the energy difference is so small
that it can only be viewed as a tendency towards closed-packed systems.
A statement about the preferred phase is not possible.}\\
$^b$ {Including zero point motion.}
\end{table}

\begin{table}
\begin{center}
\caption{\label{surfsummtable}Surface energies of (100), (110), and
(111) surface, in units of$\frac{mE_h}{a_{bohr}^2}$; 
1 $\frac{mE_h}{a_{bohr}^2}$=1.5567 $\frac{J}{m^{2}}$ }
\vspace{5mm}
\begin{tabular}{ccccccc}
surface & LDA (3.37 \AA) & \multicolumn{3}{c}{PWGGA (3.44 \AA)} & 
Ref. \onlinecite{Kokkoetal} (LDA, 3 layers, at 3.41 \AA)   \\
& $[3s2p]$ & $[3s2p]$ & $[4s3p]$ &  $[4s3p1d]$ &    \\
(100) & 0.41 & 0.37 & 0.30 & 0.30 & 0.33 \\
(110) & 0.42  & 0.37 & 0.32 & 0.32 & 0.35 \\
(111) & 0.49 & 0.44 & 0.34 & 0.36 & 0.40 \\
\end{tabular}
\end{center}
\end{table}

\end{document}